\documentclass[conference,a4paper]{IEEEtran}
\IEEEoverridecommandlockouts

\usepackage{cite}
\usepackage{amsmath,amssymb,amsfonts}
\usepackage{algorithmic}
\usepackage{graphicx}
\usepackage{textcomp}
\usepackage{xcolor}
\usepackage{subcaption}   % modern subfigure support

\usepackage{svg}
\usepackage{pdfpages}
\usepackage{tikz}
\usepackage{tabularx}
\usepackage{booktabs}
\usepackage{algorithm}
\usepackage{placeins}
\usepackage{float}
\usepackage{setspace}
\usepackage[top=0.75in, bottom=1.0in, left=0.625in, right=0.625in]{geometry}
\usepackage{rotating}

\usepackage{lipsum}

\def\BibTeX{{\rm B\kern-.05em{\sc i\kern-.025em b}\kern-.08em
    T\kern-.1667em\lower.7ex\hbox{E}\kern-.125emX}}

\begin{document}

\title{Impact of RIS Orientation on Throughput in UAV-Assisted Wireless Systems}

\author{
\IEEEauthorblockN{
Zawar Hussain$^{1,2}$,
Faran Awais Butt$^{3}$,
Ali Hussein Muqaibel$^{3,4}$,
Saleh Ahmed Alawsh$^{3}$,
Ijaz Haider Naqvi$^{1}$
}
\\
\IEEEauthorblockA{
$^{1}$Department of Electrical Engineering, Lahore University of Management Sciences (LUMS), Lahore, Pakistan \\
$^{2}$School of Engineering, University of Management \& Technology (UMT), Lahore, Pakistan \\
$^{3}$Center for Communication Systems and Sensing, King Fahd University of Petroleum \& Minerals, Dhahran, Saudi Arabia \\
$^{4}$Electrical Engineering Department, King Fahd University of Petroleum \& Minerals, Dhahran, Saudi Arabia \\
\{zawar.hussain, ijaznaqvi\}@lums.edu.pk, \{faranawais.butt, muqaibel, salawsh\}@kfupm.edu.sa}}

%\thanks{Corresponding Author: Faran Awais Butt (email: faranawais.butt@kfupm.edu.sa)}

\IEEEoverridecommandlockouts
\IEEEoverridecommandlockouts\IEEEpubid{\makebox[\columnwidth]{979-8-3315-9878-5/25/\$31.00~\copyright~2025} \hspace{\columnsep}\makebox[\columnwidth]{ }}

\maketitle

\begin{abstract}
This paper investigates the impact of Reconfigurable Intelligent Surface (RIS) orientation on the throughput performance of Unmanned Aerial Vehicle (UAV)-assisted wireless communication systems. Specifically, we study how physical rotation of the RIS, through controlled azimuth and elevation adjustments, influences the effective channel and data rate. A UAV-mounted RIS enables directional alignment to serve ground users in scenarios where the direct Base Station (BS)-to-user path is blocked. Using the SimRIS channel simulator, we analyze the system under various rotation angles and present performance heatmaps that highlight optimal RIS orientations. The study shows that RIS alignment has a substantial effect on achievable rates, thereby motivating orientation-aware optimization in practical deployments.  \\
\end{abstract}


\section{Introduction}
With the exponential growth in users and Internet of Things (IoT) devices, there is an ever-increasing need for innovative communication solutions, some of which were addressed in fifth-generation (5G) mobile communications \cite{b1}. The incorporation of Multi-input Multi-output (MIMO) and millimeter (mm)-wave communication in 5G, however, grapples with issues such as limited control over the wireless channel and substantial power consumption of the wireless interface \cite{b2}.

Reconfigurable Intelligent Surfaces (RIS), also known as Intelligent Reflecting Surfaces (IRS) or Reconfigurable Intelligent Meta-surfaces (RIM), have been introduced as a solution. These surfaces are equipped with numerous antennas on a dielectric substrate, enabling phase and amplitude adjustments of incident signals \cite{b3},\cite{b4}. The control circuit board, driven by an FPGA or microcontroller, manipulates the reflection amplitude and phase of RIS elements, while a copper backplane mitigates signal leakage \cite{b1}.

The versatility of these meta-surfaces, especially their ability to offer phase shifts and modify polarization using active elements such as diodes, facilitates the steering of electromagnetic waves in desired directions through software-controlled methods. With each RIS encompassing $N$ reflecting elements, the data rate improves proportionally as the number of elements increases. Key objectives of RIS technology include extending coverage, enhancing physical layer security, supporting large-scale device-to-device communication, and enabling wireless information and power transfer \cite{b5}. Since RIS panels typically consist of reflecting elements without dedicated RF chains for transmission and reception, they maintain a compact footprint, consume very low power, and support environmentally friendly deployment. As an enabling technology, RIS also introduces a resource-allocation dimension where energy efficiency must be jointly considered alongside throughput in UAV-assisted deployments \cite{b16}.

RIS has shown potential to transform wireless communications by offering unprecedented control over the propagation environment \cite{b6}. However, in conventional fixed installations, such as RIS panels mounted on building exteriors, the coverage is often limited to a semicircular 180$^{\circ}$ field. To address this, recent work proposes mounting RIS on mobile aerial platforms like UAVs and tethered balloons, allowing enhanced coverage and better handling of blockages in terrestrial networks. Related advances in RIS-inspired radar waveform design also emphasize bandwidth compression with controlled resolution trade-offs.

This work investigates the impact of RIS orientation when deployed on aerial platforms such as UAVs. By leveraging the freedom of 180$^{\circ}$ physical rotation, we study how adjusting the azimuth and elevation angles of RIS can influence system performance and throughput. A key challenge is determining the optimal rotation angles that maximize performance while minimizing deviation from initial alignment. We evaluate this using the SimRIS Channel Simulator, a MATLAB-based tool that supports channel modeling for RIS-assisted systems \cite{b7}. The simulator takes into account RIS orientation, environmental features, and propagation characteristics such as Line-of-Sight (LOS) probability, array responses, and other parameters.
%%%%%%%%%%%%%%%
\begin{figure*}
\centering
\includegraphics[width=2.1\columnwidth]{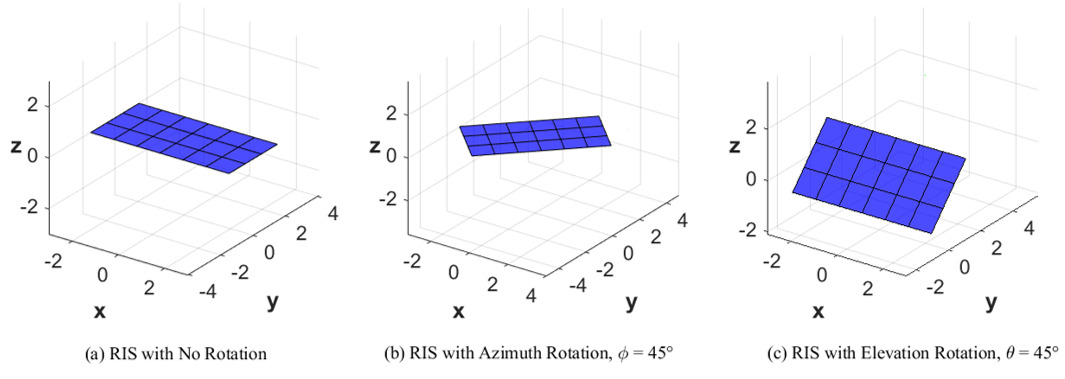}
%\vspace{-0.5in}
\caption{Illustration of RIS Orientation. }
\label{figa}
\end{figure*}

%%%%%%%%%%%%

The key contributions of this paper are summarized below:

\begin{itemize}
\item The impact of rotatable RIS mounted on UAV platforms is studied, highlighting throughput enhancement through optimized azimuth and elevation orientations.
\item Performance trends are analyzed using 3D heat maps, with identification of angle regions that maximize achievable rate.
%\item An optimization problem is formulated and solved to balance achievable rate with rotation constraints, providing a practical method to select RIS orientation under system limitations.
\end{itemize}
%\vspace{-0.25in}
The rest of the paper is structured as follows. Section~II introduces the system model, while Section~III details the simulation setup and presents the performance evaluation. Finally, Section~IV concludes the paper with key findings and insights.

\section{System Model}
\begin{figure}
\centerline{\includegraphics[width=9cm, height=6cm]{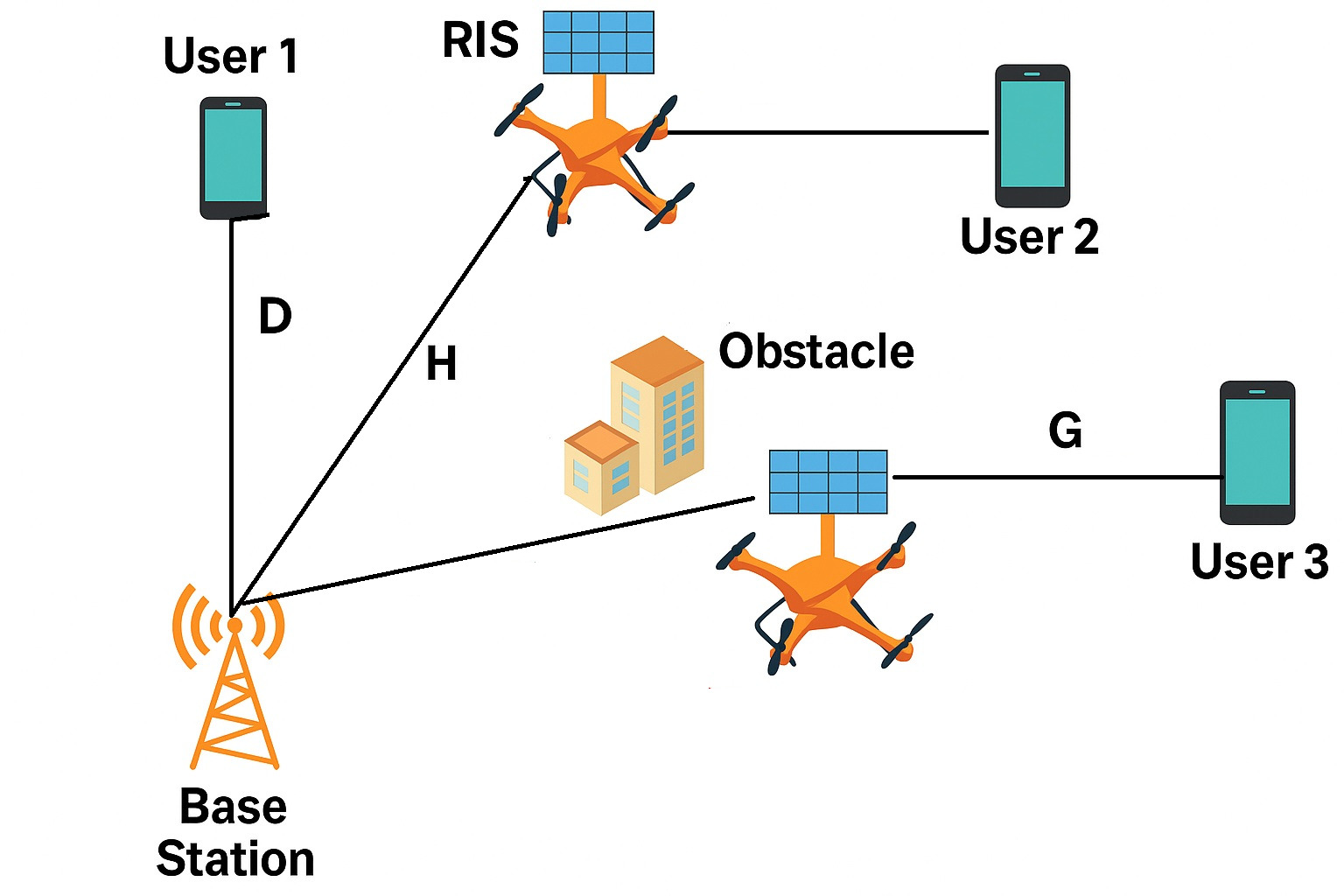}}
\caption{Illustration of UAV-mounted RIS-assisted wireless communication system. }
\label{figb}
\end{figure}

%\vspace{-0.25in}
Fig.~\ref{figa} illustrates the effect of azimuth and elevation rotations on a typical RIS panel in a 3D geometry. In the reference case with no rotation, the RIS is aligned with the global coordinate axes. Azimuth rotation corresponds to turning the RIS panel around the vertical $z$-axis by an angle $\phi$, whereas elevation rotation corresponds to tilting the panel around the horizontal $x$-axis by an angle $\theta$. 
We consider a UAV-mounted RIS-assisted MIMO wireless communication system, where multiple UAVs are equipped with RIS to enhance connectivity between a base station (BS) and ground users, as shown in Fig.~\ref{figb}. In scenarios where the direct BS-to-user link experiences blockage, the RIS assists by reflecting signals toward the intended receiver.

% \begin{figure}[t!]
%     \centering
%     \begin{subfigure}[b]{0.3\textwidth}
%         \includegraphics[width=\textwidth]{rota.eps}
%         \caption{RIS with No Rotation}
%     \end{subfigure}
%     \hfill
%     \begin{subfigure}[b]{0.3\textwidth}
%         \includegraphics[width=\textwidth]{rotb.eps}
%         \caption{RIS with Azimuth Rotation, $\phi=45^\circ$}
%     \end{subfigure}
%     \hfill
%     \begin{subfigure}[b]{0.3\textwidth}
%         \includegraphics[width=\textwidth]{rotc.eps}
%         \caption{RIS with Elevation Rotation, $\theta=45^\circ$}
%     \end{subfigure}
%     \caption{Illustration of RIS Orientation}
%     \label{fig:ris_orientation}
% \end{figure}

%\vspace{-0.25in}

%\vspace{-0.25in}

% \begin{figure}[t] % [t] = top of page
%     \centering
%     % three subfigures each about 0.3\textwidth so they fit in one row
%     \begin{subfigure}[b]{0.16\textwidth}
%         \includegraphics[width=\linewidth]{rota.eps}
%         \caption{RIS with No Rotation}
%     \end{subfigure}\hfill
%     \begin{subfigure}[b]{0.16\textwidth}
%         \includegraphics[width=\linewidth]{rotb.eps}
%         \caption{RIS with Azimuth Rotation, $\phi=45^\circ$}
%     \end{subfigure}\hfill
%     \begin{subfigure}[b]{0.16\textwidth}
%         \includegraphics[width=\linewidth]{rotc.eps}
%         \caption{RIS with Elevation Rotation, $\theta=45^\circ$}
%     \end{subfigure}
%     \caption{Illustration of RIS Orientation}
%     \label{rott}
% \end{figure}

To improve link quality, the RIS is assumed to be capable of mechanical rotation along its azimuth and elevation axes. Two control strategies are possible: (i) rotating the entire UAV platform to align the RIS, and (ii) rotating only the RIS panel while keeping the UAV fixed. In this work, we adopt the second strategy, which offers lower energy consumption and simpler UAV control. However, it comes with its own tradeoffs such as the need for precise low-power motors and control circuits to manage RIS orientation independently and maintain alignment under UAV motion or wind. Furthermore, accurate RIS alignment benefits from localization methods that employ multimodal calibration and optimization \cite{b14}.

% \begin{figure}[t] % [t] = top of page
%     \centering
%     % three subfigures each about 0.3\textwidth so they fit in one row
%     \begin{subfigure}[b]{0.16\textwidth}
%         \includegraphics[width=\linewidth]{rota.eps}
%         \caption{RIS with No Rotation}
%     \end{subfigure}\hfill
%     \begin{subfigure}[b]{0.16\textwidth}
%         \includegraphics[width=\linewidth]{rotb.eps}
%         \caption{RIS with Azimuth Rotation, $\phi=45^\circ$}
%     \end{subfigure}\hfill
%     \begin{subfigure}[b]{0.16\textwidth}
%         \includegraphics[width=\linewidth]{rotc.eps}
%         \caption{RIS with Elevation Rotation, $\theta=45^\circ$}
%     \end{subfigure}
%     \caption{Illustration of RIS Orientation}
%     \label{rott}
% \end{figure}

% \begin{figure}[t]
%     \centering
%     \begin{subfigure}[b]{0.16\textwidth}
%         \includegraphics[width=1\textwidth]{rota.eps}
%         \caption{RIS with No Rotation}
%     \end{subfigure}
%     \begin{subfigure}[b]{0.16\textwidth}
%         \includegraphics[width=1\textwidth]{rotb.eps}
%         \caption{RIS with Azimuth Rotation, $\phi=45^\circ$}
%     \end{subfigure}
%     \begin{subfigure}[b]{0.16\textwidth}
%         \includegraphics[width=1\textwidth]{rotc.eps}
%         \caption{RIS with Elevation Rotation, $\theta=45^\circ$}
%     \end{subfigure}
%     \caption{ Illustration of RIS Orientation}
%     \label{rott}
% \end{figure}

These mechanical rotations alter the orientation of the individual RIS elements, thereby changing the effective reflection direction of the incident wave. To accurately incorporate these effects into the channel model, the RIS element orientations are updated using standard 3D rotation matrices. Specifically, the overall rotation matrix $\mathbf{R}$ is obtained from successive rotations around the $z$-axis (azimuth) and the $x$-axis (elevation), as defined below.

\begin{equation}
\mathbf{R}_z(\phi) = 
\begin{bmatrix}
\cos(\phi) & -\sin(\phi) & 0 \\
\sin(\phi) & \cos(\phi) & 0 \\
0 & 0 & 1
\end{bmatrix}
\end{equation}

\begin{equation}
\mathbf{R}_x(\theta) = 
\begin{bmatrix}
1 & 0 & 0 \\
0 & \cos(\theta) & -\sin(\theta) \\
0 & \sin(\theta) & \cos(\theta)
\end{bmatrix}
\label{rotation_matrices}
\end{equation}

These transformations allow the RIS reflection direction to be dynamically adjusted during deployment.
\begin{figure}
\centerline{\includegraphics[width=9cm, height=5.8cm]{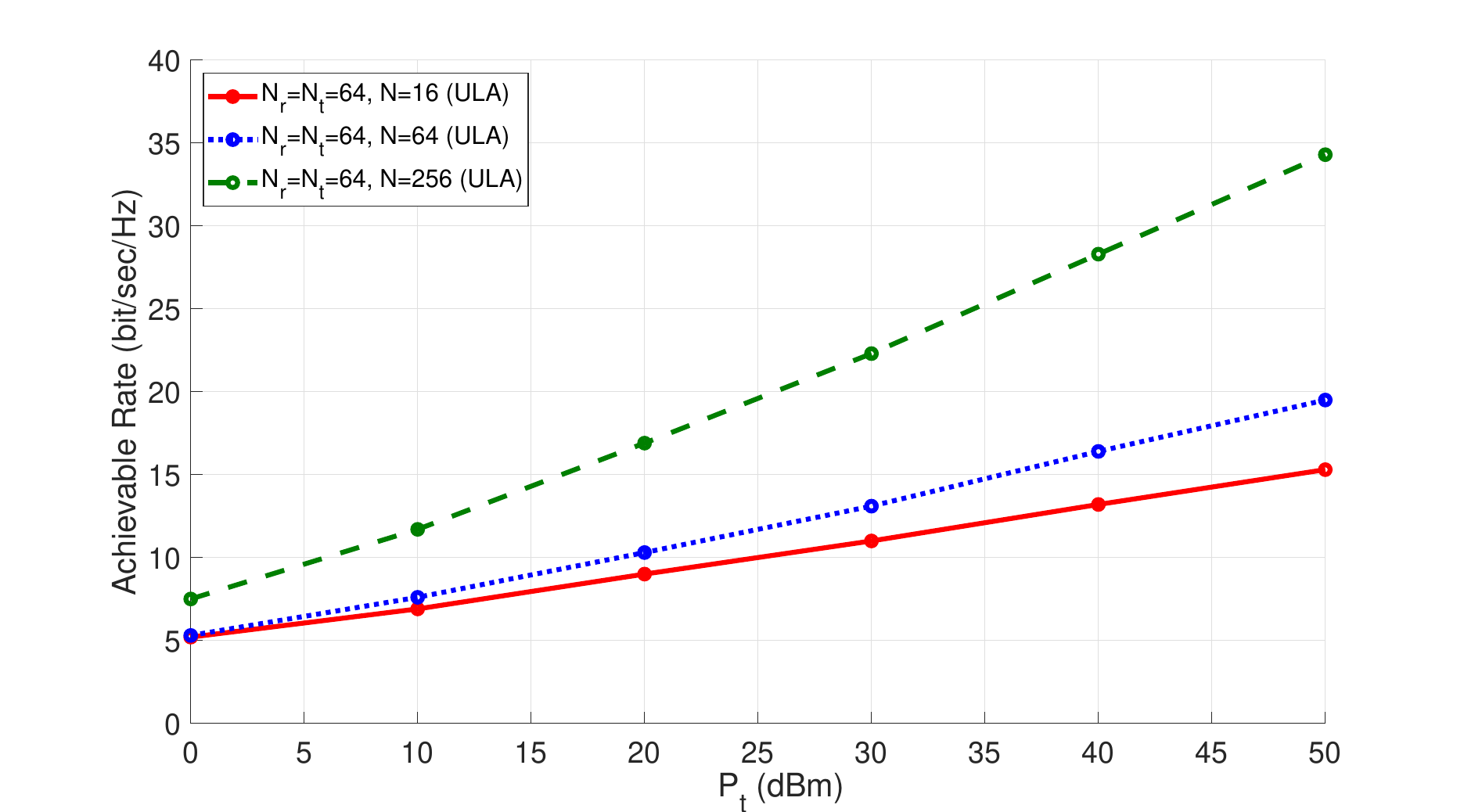}}
\caption{Achievable rate vs. transmit power $P_t$ for different RIS sizes ($N$).}
\label{z3}
\end{figure}

The system employs a MIMO architecture. Let $N_t$ denote the number of transmit antennas at the base station (BS), $N_r$ denote the number of receive antennas at the user terminal, $N$ denote the number of reflecting elements on the RIS. The end-to-end equivalent channel matrix $\mathbf{C} \in \mathbb{C}^{N_r \times N_t}$ is expressed as

\begin{equation}
\mathbf{C} = \mathbf{G} \boldsymbol{\Phi} \mathbf{H} + \mathbf{D}
\label{eq:main_equation}
\end{equation}

\noindent
where $\mathbf{G} \in \mathbb{C}^{N_r \times N}$ represents the channel coefficients matrix from RIS to user,  
$\boldsymbol{\Phi} = \operatorname{diag}(e^{j\varphi_1}, \ldots, e^{j\varphi_N})$ is the RIS diagonal phase shift matrix,  
$\mathbf{H} \in \mathbb{C}^{N \times N_t}$ represents the channel coefficients matrix from BS to RIS, $\mathbf{D} \in \mathbb{C}^{N_r \times N_t}$ denotes the direct channel between BS and user.

\section{Simulation Results}

Simulations were conducted in both indoor and outdoor settings, with the system operating at a frequency of 28~GHz. 
The transmitter, $T_{x}$, is located at the coordinates $(0,25,2)$ in the $(x,y,z)$ plane, while the receiver, $R_{x}$, is positioned at $(45,45,1)$. 
The RIS is placed at $(40,50,2)$, with its reflecting surface oriented towards both the transmitter and the receiver to ensure effective signal reflection.
Two different RIS deployment scenarios are considered in the simulations, namely indoor and outdoor.  For indoor, the transmitter height is limited to 3 meters, the receiver must be below 2 meters, and the RIS–receiver distance is constrained to a maximum of 10 meters. In contrast, the outdoor setup allows transmitter heights up to 20 meters and does not restrict RIS–receiver distance. Both environments support mmWave frequencies such as 28 GHz and 73 GHz, but use different channel models, dense multipath for indoor and 3GPP based path loss for outdoor settings. These parameters directly affect the simulation geometry, validation checks, and channel behavior in the SimRIS GUI \cite{b15}\cite{b12}. Two antenna array configurations are examined at the transmitter and receiver: the uniform linear array (ULA) and the uniform planar array (UPA), while the RIS is consistently modeled as a uniform planar array.
The ergodic achievable rate for the MIMO system is evaluated, which is defined by the following equation:

\begin{equation}
R = \log_2 \left(\text{det} \left(I_{N_r} + \frac{P_t}{\sigma^2}\mathbf{C}\mathbf{C^H}\right)\right) \text{ bits/sec/Hz}
\end{equation}
where $I_{N_r}$ is the $N_r \times N_r$ identity matrix, 
$\sigma^2$ denotes the average noise power, 
and $P_{t}$ represents the transmit power.

\begin{figure}[t]
\includegraphics[width=\columnwidth]{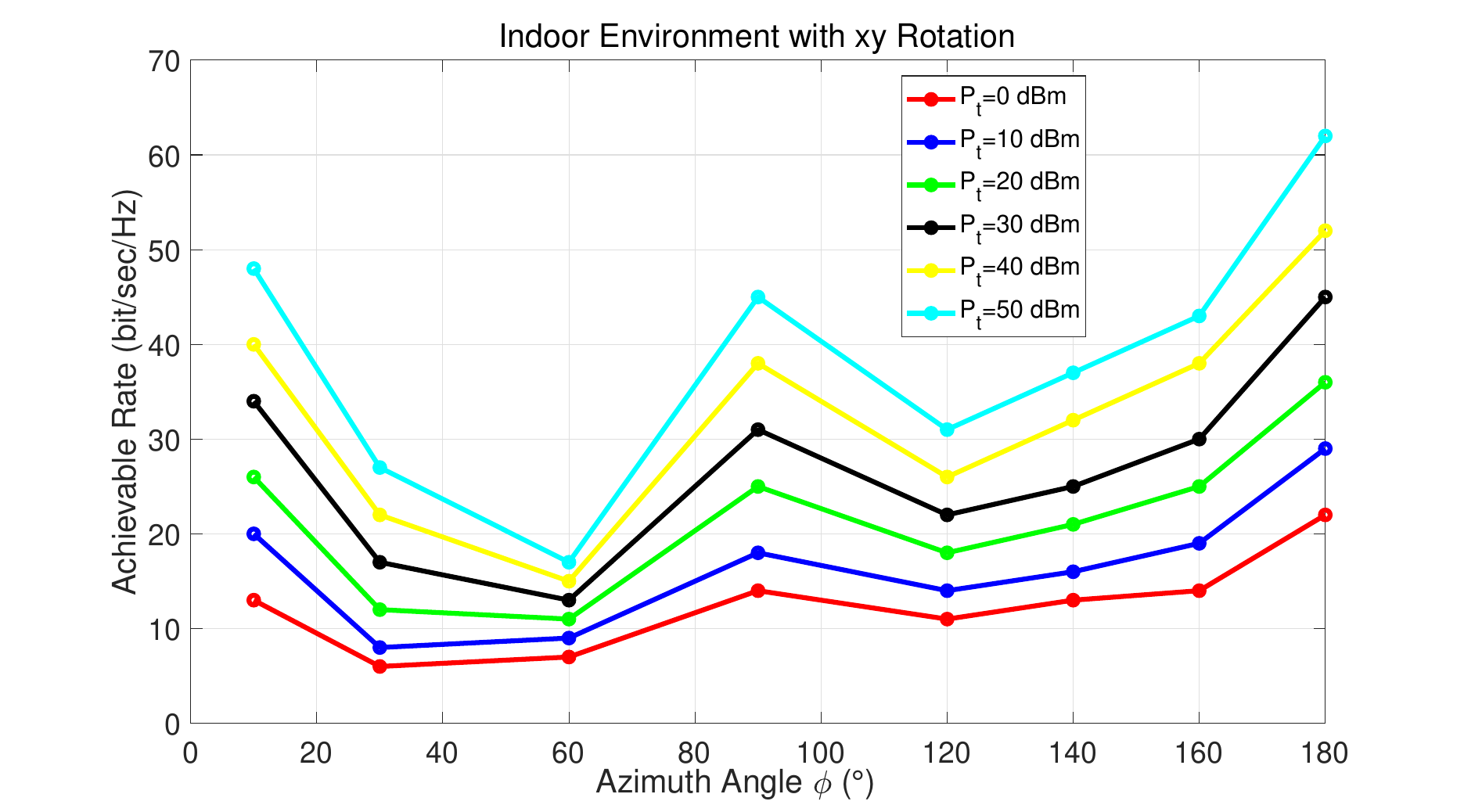}
\caption{Achievable rate with azimuth rotation
under varying $P_{t}$}
\label{outdoor}
\end{figure}

\begin{figure}[t!]
\includegraphics[width=\columnwidth]{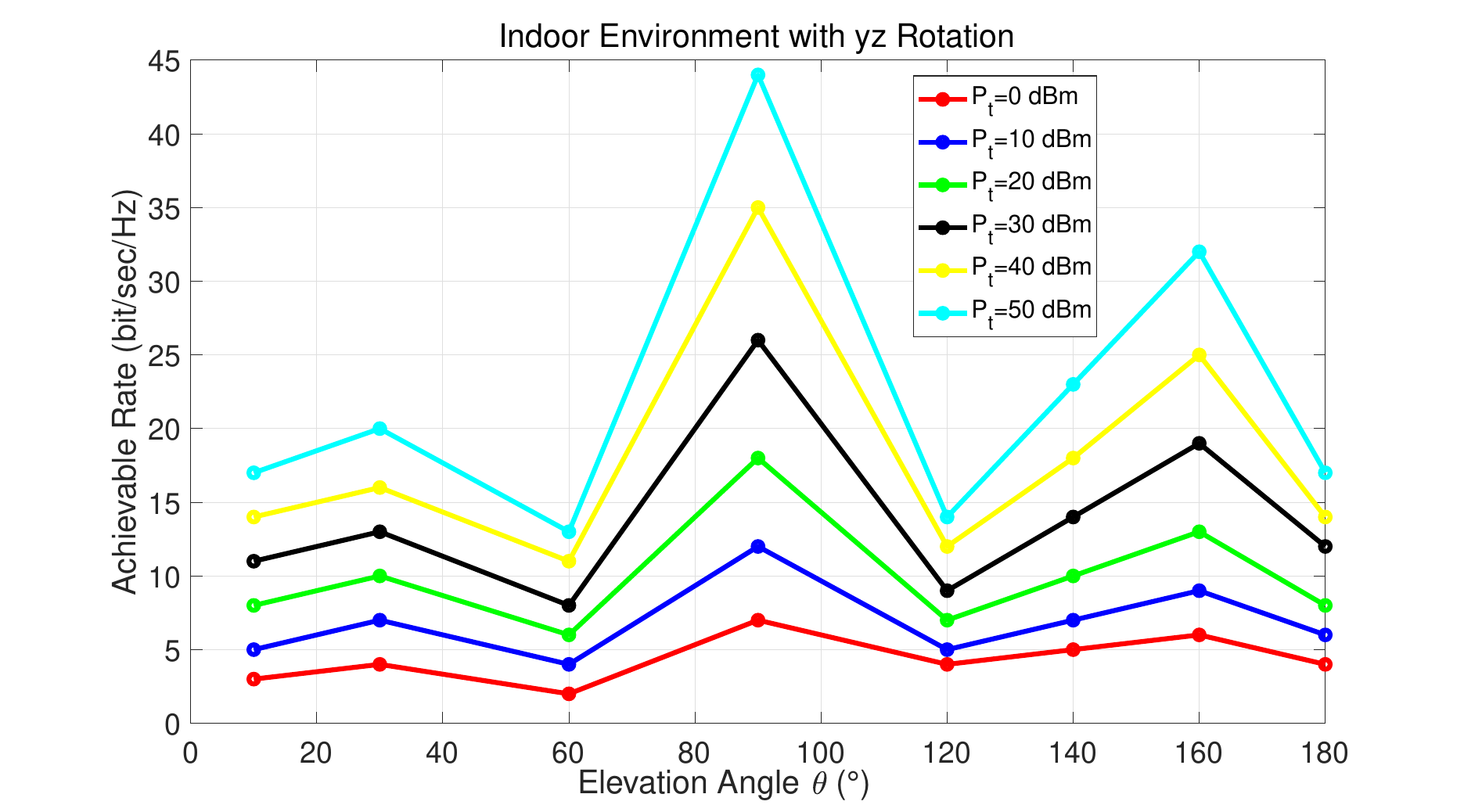}
\caption{Achievable rate with elevation rotation
under varying $P_{t}$}
\label{fig5}
\end{figure}

\subsection{Achievable Rate with no Rotation Effect }\label{AA}

Fig.~\ref{z3} illustrates the impact of varying the number of RIS elements while keeping the number of transmit and receive antennas fixed. The simulation was conducted without any rotation, with both azimuth and elevation angles set to zero degrees in the indoor environment. The achievable rate was observed to increase with higher transmit power \( P_t \). A significant improvement in data rate was achieved as both \( P_t \) and the number of RIS elements increased. In particular, a peak rate exceeding 35~bits/sec/Hz was recorded when the largest RIS configuration was used at a transmit power of 50~dBm.

\begin{figure*}[t]
\centering

\begin{subfigure}[t]{0.48\textwidth}
    \centering
    \includegraphics[width=\linewidth]{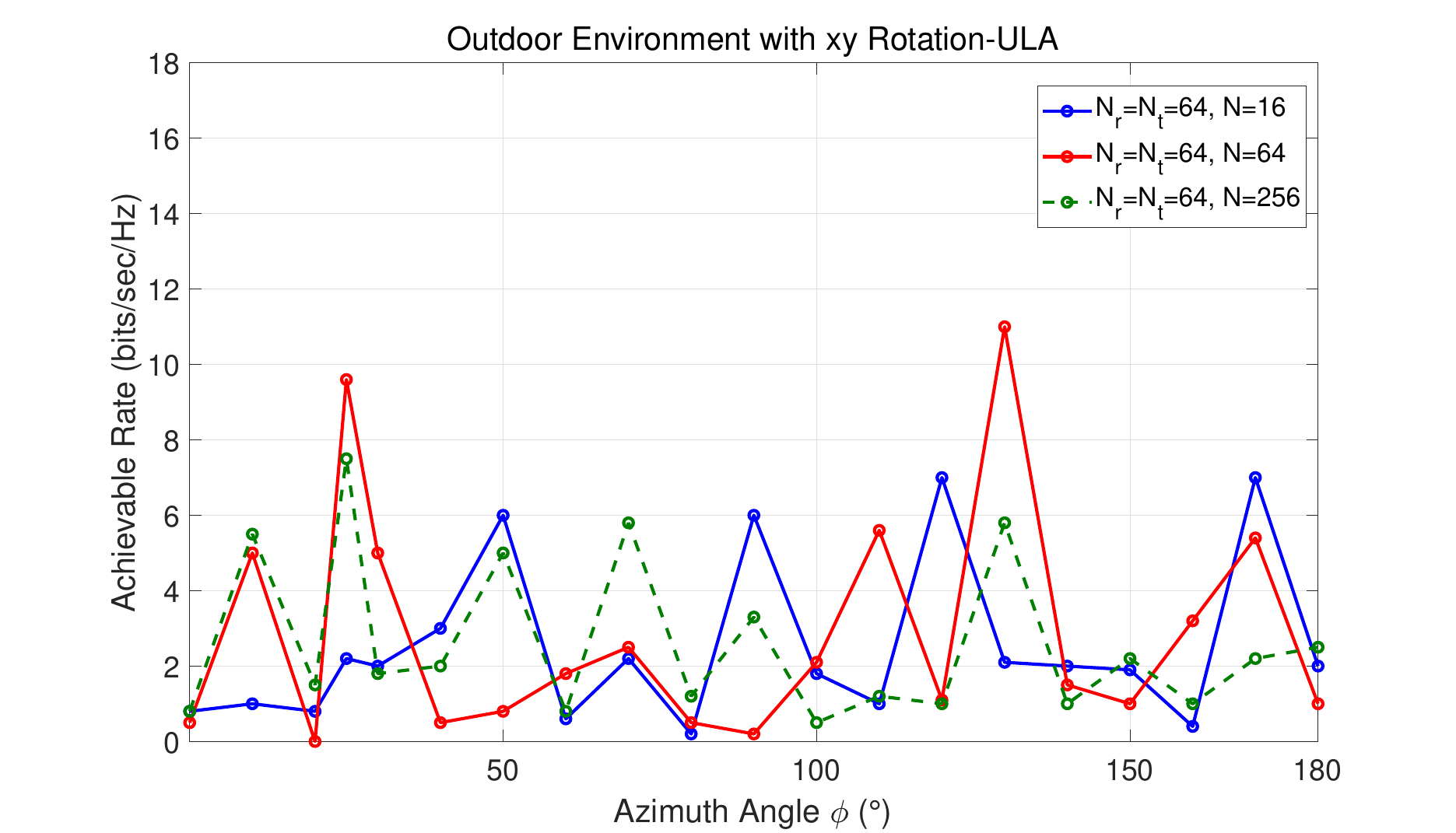}
    \caption{ULA, Azimuth Rotation}
\end{subfigure}
\hfill
\begin{subfigure}[t]{0.48\textwidth}
    \centering
    \includegraphics[width=\linewidth]{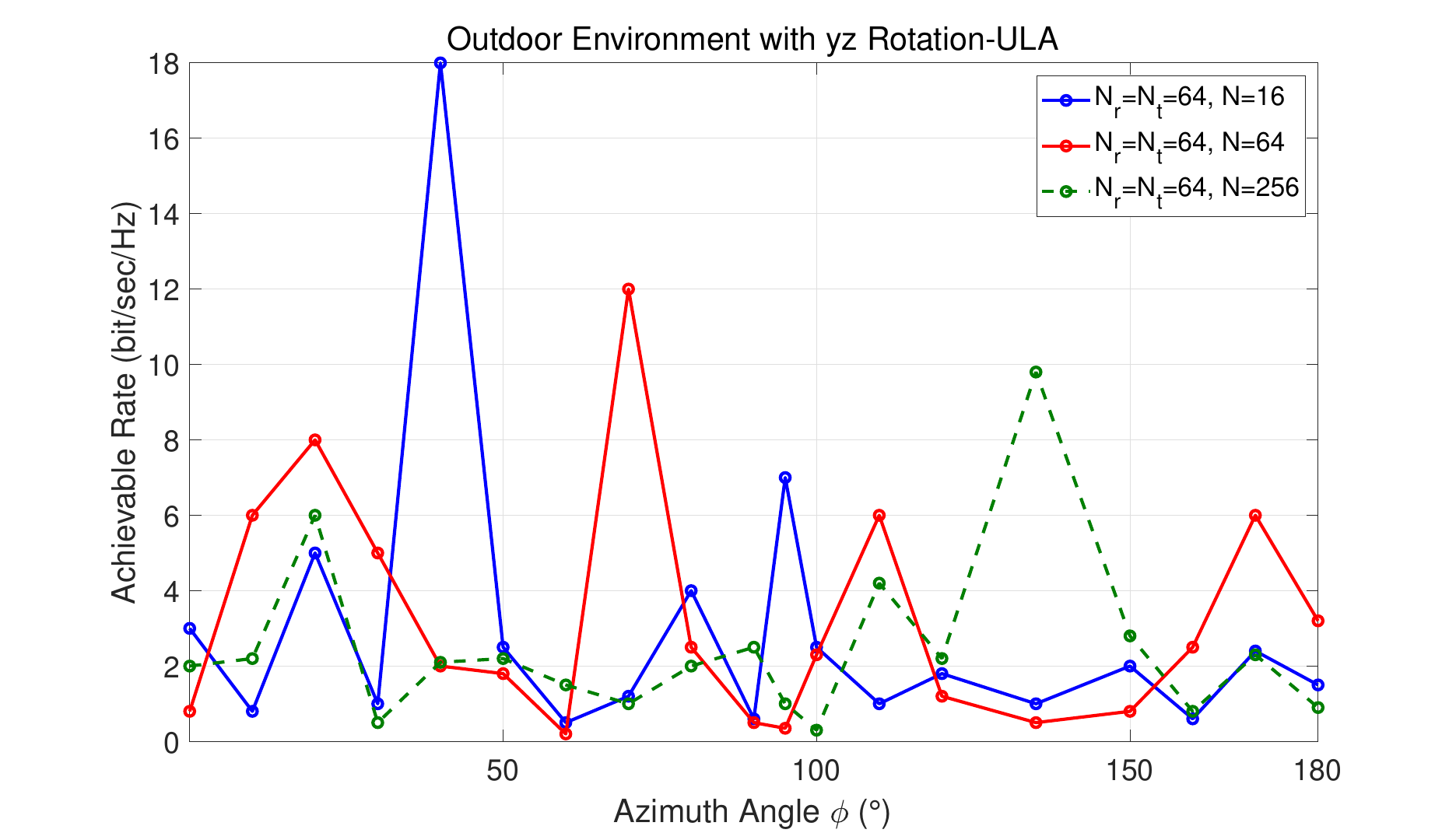}
    \caption{ULA, Elevation Rotation}
\end{subfigure}

\begin{subfigure}[t]{0.48\textwidth}
    \centering
    \includegraphics[width=\linewidth]{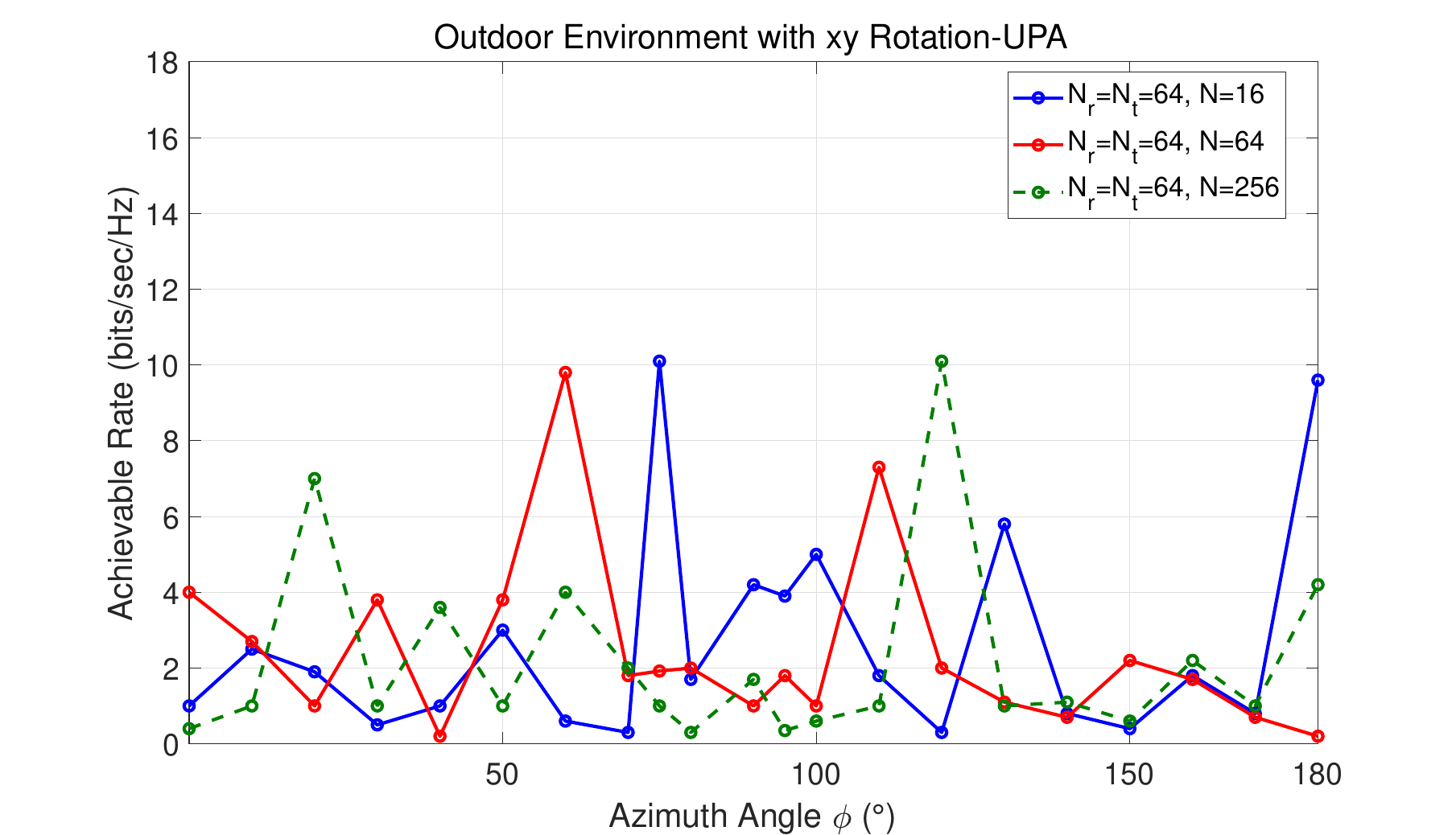}
    \caption{UPA, Azimuth Rotation}
\end{subfigure}
\hfill
\begin{subfigure}[t]{0.48\textwidth}
    \centering
    \includegraphics[width=\linewidth]{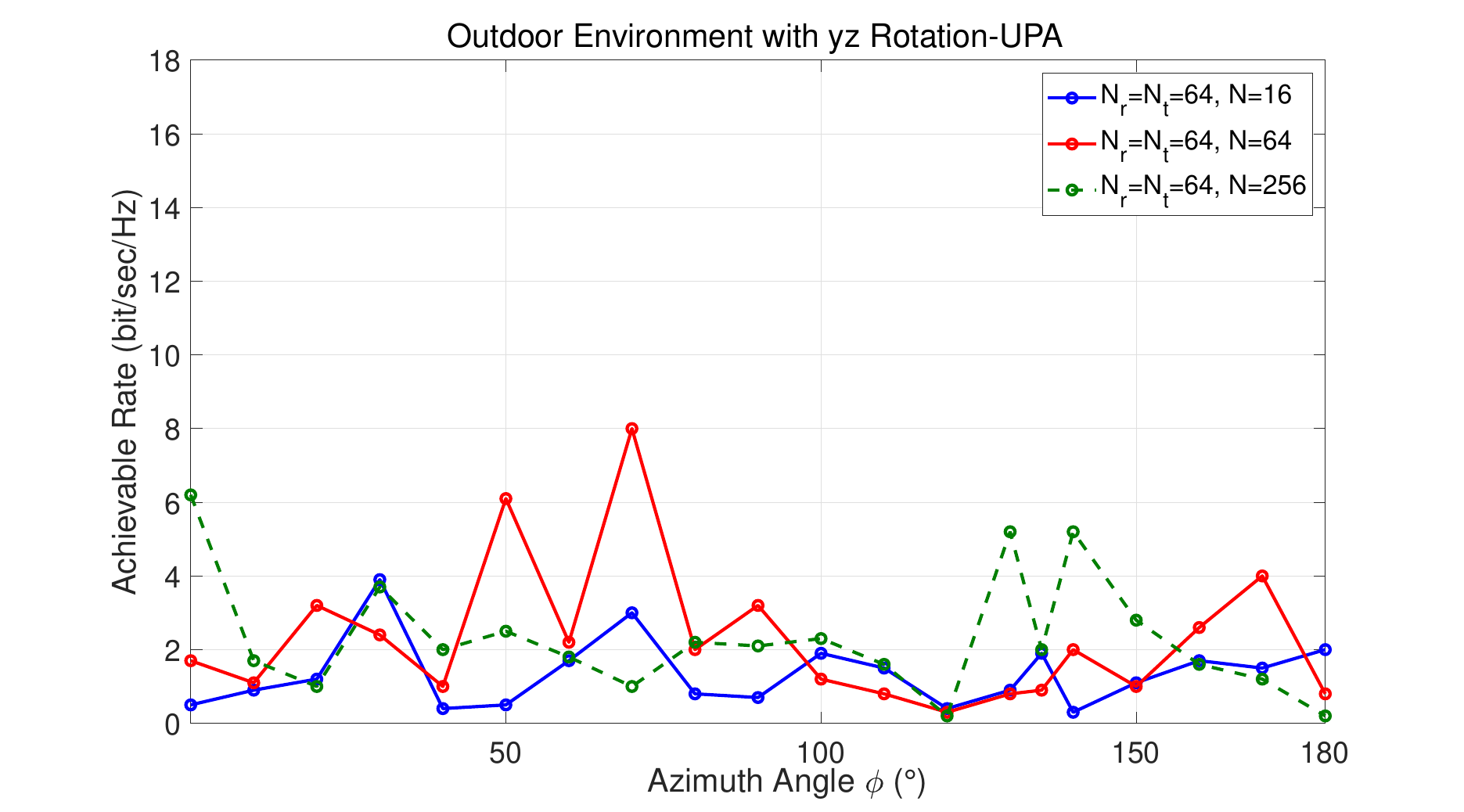}
    \caption{UPA, Elevation Rotation}
\end{subfigure}

\caption{Achievable rate under different RIS rotations and configurations in outdoor ULA/UPA setups.}
\label{fig6}
\end{figure*}

%%%%%%%%%%%%%%%

\begin{figure*}[t]
\centering
\includegraphics[width=\textwidth, height= 9cm]{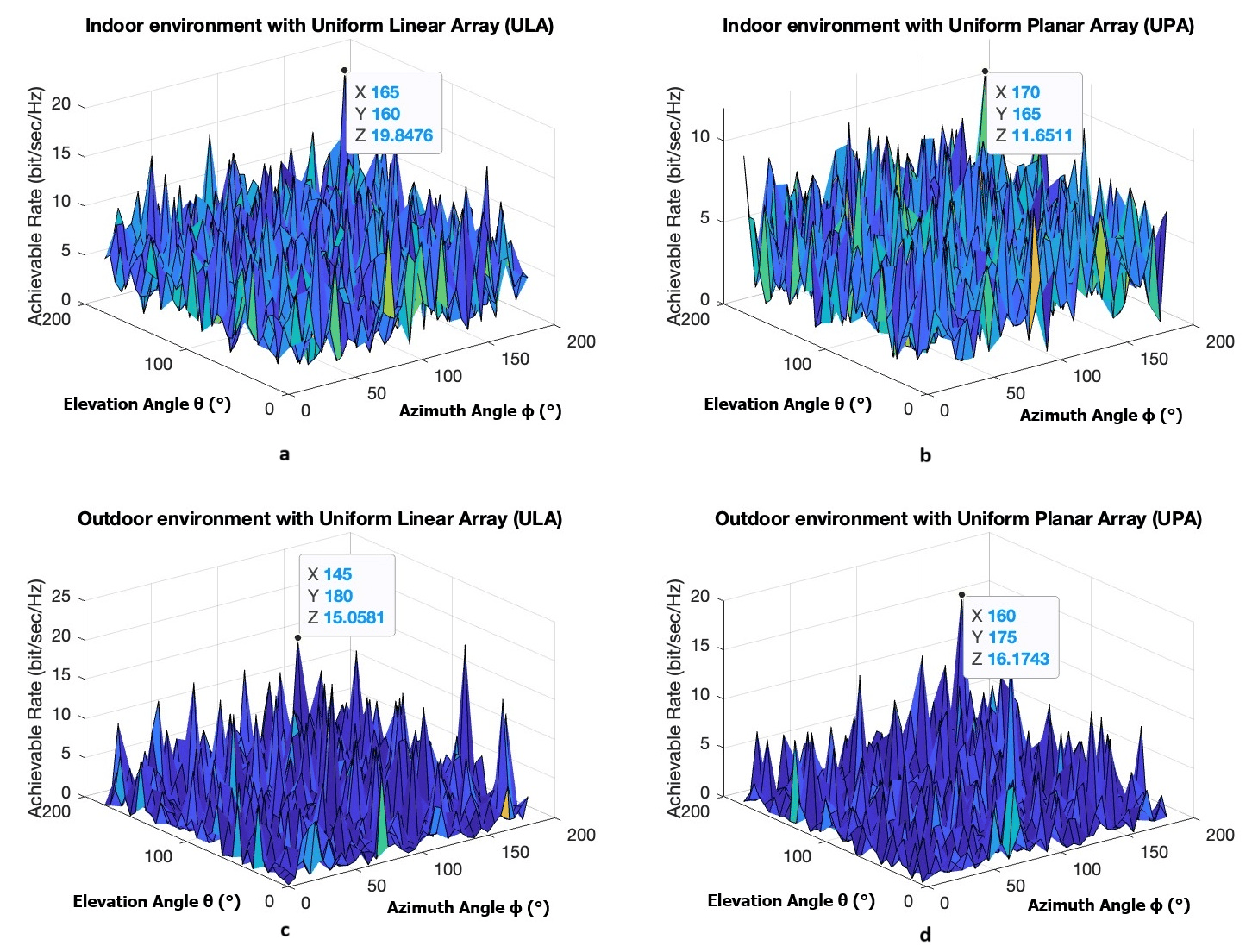}
\caption{Achievable rate with simultaneous azimuth ($\phi$) and elevation ($\theta$) rotation for ULA and UPA configurations}
\label{SAE}
\end{figure*}
% \vspace{-0.23in}
\subsection{Achievable Rate with Rotation Effect }\label{AA}

%%%%%%%%%%%%
In this scenario, we investigate the effect of RIS rotation on the achievable rate by varying both elevation and azimuth angles. 
To ensure clarity, the modeled outdoor setup is defined as follows: the BS, $T_x$, receiver, $R_x$, and UAV-mounted RIS are placed in a Cartesian $(x,y,z)$ coordinate system, with the RIS surface oriented towards both transmitter and receiver to enable effective reflection. 
Fig.~\ref{outdoor} illustrates how the achievable rate varies with azimuth angle when the RIS undergoes mechanical rotation in the xy-plane for different transmit powers ranging from 0 dBm to 50 dBm. The results clearly show that RIS orientation has a strong impact on system throughput. At certain azimuth angles, constructive alignment of reflected paths yields significantly higher rates, whereas at others, misalignment reduces performance. Increasing the transmit power uniformly shifts the achievable rate upward, but the relative fluctuations across angles remain, confirming that RIS rotation, not just power scaling, is a critical factor in maximizing spectral efficiency.

This behavior highlights the beam-steering capability of a rotatable RIS: by adjusting its azimuth angle, the RIS can dynamically align with favorable propagation directions, mitigating non-line-of-sight. The sharp variations in rate across angles emphasize the necessity of joint optimization of RIS orientation and transmit power, ensuring robust coverage and improved capacity in practical deployments. In SimRIS, the indoor and outdoor scenarios are differentiated by specific geometric and propagation constraints. In Fig.~\ref{fig5} the impact of transmit power on the achievable rate across different RIS rotation angles is presented in an indoor scenario with $N_r,N_t$=64 and $N$=256. 
Each curve corresponds to a specific $P_t$ value, showing how the system benefits from both higher transmit power and optimal RIS alignment. 
The results reveal that while power scaling naturally improves throughput, the achievable rate remains highly sensitive to RIS orientation. 
Notably, the steep peaks observed at certain angles indicate constructive alignment of multipath components, reinforcing the need to co-optimize both power and angular alignment.

Fig.~\ref{fig6} illustrates the achievable rate variation in outdoor environments under RIS rotations along azimuth and elevation angles, considering different RIS sizes and both ULA and UPA array configurations. The results show that the achievable rate exhibits strong fluctuations across rotation angles, both in terms of frequency and amplitude, highlighting the critical role of precise RIS orientation. In the ULA cases (Fig.~\ref{fig6}(a) and (b)), the fluctuations are more frequent and larger in amplitude due to the narrower angular coverage of linear arrays, making them highly sensitive to orientation changes. By contrast, UPA cases (Fig.~\ref{fig6}(c) and (d)) demonstrate fluctuations with smaller amplitudes, as planar arrays provide wider angular coverage and thus greater robustness against misalignment. Also, increasing the number of reflecting elements $N$ improves peak achievable rates but also amplifies sensitivity to angular deviations, since larger RIS panels enhance constructive alignment while deepening misalignment nulls. 
%The results indicate that RIS orientation, array type, and panel size jointly shape the achievable performance in outdoor deployments, highlighting the importance of precise rotation control and adaptive alignment mechanisms for reliable UAV-assisted RIS systems.

To capture the joint influence of RIS orientation, Fig.~\ref{SAE} presents 3D heat maps analysis showing the achievable rate variation with simultaneous azimuth and elevation adjustments under a fixed transmit power of 10 dBm in both indoor and outdoor scenarios. Analysis of achievable rate is carried out with ULA and UPA antenna setups with $N_r, N_t$ =64 and $N$=256. The highest achievable rates are achieved at $\phi$ ranging from $165^{o}$ to $180^{o}$ and $\theta$ ranging from $145^{o}$ to $170^{o}$. These findings highlight the significance of precise RIS orientation to maximize throughput. These heatmaps clearly reveal that performance is highly sensitive to orientation, with well-defined peaks indicating favorable alignment zones. For instance, in the indoor ULA case in Fig.~\ref{SAE}(a) , rates peak at $\phi \approx 165^\circ$ and $\theta \approx 160^\circ$, reaching values close to 20 bits/sec/Hz. In contrast, outdoor UPA scenario in Fig.~\ref{SAE}(d) experiences more fluctuation and attenuation, yet still demonstrates considerable throughput gain with optimal orientation. These observations validate the critical role of simultaneous angular control and suggest that RIS rotation must be treated as a key design variable for UAV-mounted systems operating under dynamic blockage conditions. In applications involving RIS, optimizing the position  of the RIS is crucial for maximizing signal quality or data rate. 
The results indicate that RIS orientation, array type, and panel size jointly shape the achievable performance in outdoor deployments, highlighting the importance of precise rotation control and adaptive alignment mechanisms for reliable UAV-assisted RIS systems. This motivates a joint optimization of UAV orientation $(\theta,\phi)$ together with RIS reflection control to fully exploit the degrees of freedom offered by aerial platforms for throughput enhancement.

%\begin{figure*}[t]
%\centering
 %   \includegraphics[width=\textwidth, height=10cm]{Outdoor_scenario1_N.jpg}
  %  \caption{RIS rotation with array configurations}
   % \label{ac}
%\end{figure*}

%\begin{figure*}[t!]
%\centering
%\includegraphics[width=16cm, height=10cm]{rotation_N_Nt.png}
%\caption{Elevation and Azimuth angle rotation in Indoor Environment with varying $N$, $N_{t}$ and $N_{r}$}
%\label{rot_N_Nt}
%\end{figure*}

\section{Conclusion}

This paper presented a comprehensive analysis of the impact of RIS orientation on the achievable rate in UAV-assisted wireless communication systems. Using a realistic channel simulator, the effects of both azimuthal and elevation rotations were examined under various propagation conditions. The results demonstrate that precise control over RIS alignment can lead to substantial throughput gains, particularly in non-line-of-sight environments. The simulations across multiple deployment scenarios revealed consistent performance enhancements associated with specific angular configurations. These observations emphasize the critical role of mechanical rotation in practical RIS implementations and support the integration of orientation feedback into system-level optimization frameworks. The findings establish orientation-aware RIS control as a viable and effective strategy for enhancing the reliability and spectral efficiency of future wireless networks.
\section*{Acknowledgment}
Faran Awais Butt, Ali Hussein Muqaibel and Saleh Alawsh would like to acknowledge
the support provided by the Deanship of Research at King
Fahd University of Petroleum \& Minerals (KFUPM) under
the Center for Communication Systems and Sensing (Grant
INCS2510).

\end{document}